\RequirePackage{fix-cm}
\documentclass[smallextended]{svjour3}       % onecolumn (second format)
\smartqed  % flush right qed marks, e.g. at end of proof
\usepackage{graphicx}
\usepackage{multirow}
\usepackage{epsfig}
\usepackage{epstopdf}
\usepackage{amsfonts}

\journalname{}

\title{The Penrose dodecahedron and the Witting polytope are identical in $\mathbb{CP}^{3}$}

\author{Mordecai Waegell and P.K. Aravind }

\authorrunning{M.Waegell, P.K. Aravind}

\institute{M.Waegell$^{1}$,P.K.Aravind$^{2}$ \at
$^{1}$Institute for Quantum Studies, Chapman University, Orange, CA.
$^{2}$Physics Department, Worcester Polytechnic Institute, Worcester, MA 01609, U.S.A.\\
\email{waegell@chapman.edu,paravind@wpi.edu}}

\date{\today}

\begin{document}
\maketitle
\begin{abstract}
The 40 states of a spin-3/2 particle used by Zimba and Penrose to prove the Kochen-Specker and Bell theorems are shown to be identical (i.e., unitarily equivalent) in $\mathbb{CP}^{3}$ to the 40 rays that follow from the vertices of the Witting polytope. The Witting polytope also gives rise to 120 rays in $\mathbb{RP}^{7}$ that give rise to well over a billion parity proofs of the Kochen-Specker theorem. The implications of these results are discussed.

\end{abstract}

\section{\label{sec:Intro}Introduction}

Some time ago Penrose \cite{Penrose} introduced 40 states of a spin-3/2 particle related to the geometry of a dodecahedron and showed, with Zimba \cite{Zimba}, that they could be used to give proofs of the Kochen-Specker(KS) theorem \cite{KS1967} (establishing quantum contextuality) and Bell's theorem \cite{Bell} (establishing quantum nonlocality). The Penrose dodecahedron, as we will term this constellation of states, is a fascinating geometrical object whose origins are shadowy and unclear. We show here that the Penrose dodecahedron can be extracted very simply from a geometrical object known as the ``Witting polytope'', a regular complex polytope in $\mathbb{C}^{4}$ introduced by the German mathematician Alexander Witting almost a century back \cite{Coxeter}. Specifically, we show that the vertices of the Witting polytope give rise to a system of rays in the complex projective space $\mathbb{CP}^{3}$ that is unitarily equivalent to the system of rays defined by the Penrose dodecahedron. The plan of this paper is as follows. In Sec.2 we introduce the 40 states of the Penrose dodecahedron, following Penrose, and obtain expressions for their components in the standard angular momentum basis. We then show how a unitary transformation can be used to cast the Penrose rays into a simplified form in which their non-vanishing components are expressed entirely in terms of the signed cube roots of unity. In Sec.3 we introduce the Witting polytope and obtain the 40 rays in $\mathbb{CP}^{3}$ that follow from it, whereupon it will be evident that they are identical to the simplified form of the Penrose rays found in Sec.2. Finally, in Sec.4, we discuss the implications of this result.

\section{\label{sec:2} The Penrose dodecahedron}

The 40 states of the Penrose dodecahedron can be defined as follows. Twenty of the states, termed \textit{explicit rays}\footnote{A ray is an equivalence class of states whose members differ from one another only by an overall complex constant. We will be concerned mainly with rays in this paper, although at times we may exhibit them in the form of normalized states.} by Penrose, are the spin  $+1/2$ projections of a spin-3/2 particle along the twenty directions from the center of a regular dodecahedron to its vertices. Following Penrose, we will label each explicit ray by the vertex of the dodecahedron with which it is associated, with the vertices labeled as in Fig.1 (we will also use the same labels to refer to the vertices, but this should cause no confusion since it should always be clear from the context whether it is the ray or vertex that is intended). The three explicit rays associated with the vertices neighboring any vertex are mutually orthogonal, and the fourth ray that completes an orthogonal tetrad with them was termed an \textit{implicit ray} by Penrose. Again we will follow Penrose and label any implicit ray by the primed letter of the vertex surrounded by the explicit rays orthogonal to it (thus $A'$ is the implicit ray that completes a mutually orthogonal tetrad with the explicit rays $F$,$E$ and $B$). The 20 explicit and 20 implicit rays, labeled by the unprimed and primed vertices of the dodecahedron respectively, make up the 40 rays of the Penrose dodecahedron.\\

\begin{figure}[htp]
\begin{center}
\includegraphics[width=0.60\textwidth]{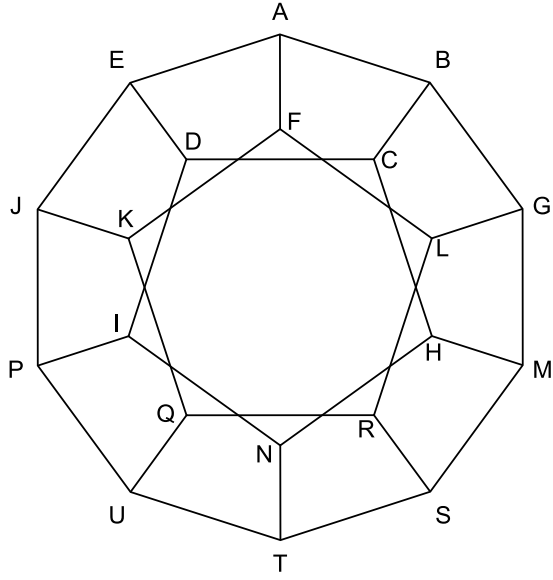}
\end{center}
\caption{A regular dodecahedron, with its vertices labeled by the letters A to U, with O omitted.}
\label{dod}
\end{figure}

In order to give explicit expressions for the Penrose rays, it is first necessary to introduce some angles. Choose a right-handed coordinate system with its origin at the center of the dodecahedron and its z-axis passing through vertex $A$ (see Fig.1), and take its positive x-axis to lie in the half-plane containing the vertices $A$ and $F$. Let $\theta_{0}=\sin^{-1}(\frac{2}{3})$, $\theta_{1}=\sin^{-1}(\frac{2\sqrt{2}}{3})$, $\phi_{1}=\sin^{-1}\bigg(\sqrt{\frac{3}{8}}\bigg)$ and $\phi_{2}=\sin^{-1}\bigg(\sqrt{\frac{3}{8}}(1+\sqrt{5})\bigg)$, with all these angles being acute. Then the polar and azimuthal angles, $(\theta,\phi)$, of the vertices of the dodecahedron are:\\

$A=(0,0),\quad F=(\theta_{0},0), \quad E=(\theta_{0},\frac{2\pi}{3}), \quad B=(\theta_{0},\frac{4\pi}{3})$ \\[1pt]

$L=(\theta_{1},\phi_{1}),\quad G=(\theta_{1},\phi_{1}+\phi_{2}), \quad C=(\theta_{1},\phi_{1}+\frac{2\pi}{3})$  \\[1pt]

$D=(\theta_{1},\phi_{1}+\phi_{2}+\frac{2\pi}{3}),\quad J=(\theta_{1},\phi_{1}+\frac{4\pi}{3}),\quad K=(\theta_{1},\phi_{1}+\phi_{2}+\frac{4\pi}{3})$ \\[1pt]

$R=(\pi-\theta_{1},\phi_{1}+\pi), \quad M=(\pi-\theta_{1},\phi_{1}+\phi_{2}+\pi ),\quad H=(\pi-\theta_{1},\phi_{1}+\frac{5\pi}{3})$ \\[1pt]

$I=(\pi-\theta_{1},\phi_{1}+\phi_{2}+\frac{5\pi}{3}), \quad P=(\pi-\theta_{1},\phi_{1}+\frac{\pi}{3}),\quad Q=(\pi-\theta_{1},\phi_{1}+\phi_{2}+\frac{\pi}{3})$ \\[1pt]

$S=(\pi-\theta_{0},\pi), \quad N=(\pi-\theta_{0},\frac{5\pi}{3}),\quad U=(\pi-\theta_{0},\frac{\pi}{3}), \quad T=(\pi,0)$ \quad (1)\\

\noindent
The vertices can also be represented by the complex numbers $\alpha=\tan(\theta/2)\exp(i\phi)$ obtained by projecting them stereographically from the unit sphere on to the equatorial plane ($\theta=\pi/2$) using the south pole ($\theta=\pi)$ as the center of projection. In the Majorana parametrization \cite{majorana}, an arbitrary (pure) state of a spin-3/2 particle is represented by three points on the unit sphere or, equivalently, by the complex numbers that correspond to them. If $\alpha_{1},\alpha_{2}$ and $\alpha_{3}$ are the complex numbers parametrizing a state, it can be represented in terms of them as \\

      $|+\frac{3}{2}\rangle + \frac{(\alpha_{1}+\alpha_{1}+\alpha_{1})}{\sqrt{3}}|+\frac{1}{2}\rangle + \frac{(\alpha_{1}\alpha_{2}+\alpha_{2}\alpha_{3}+\alpha_{3}\alpha_{1})}{\sqrt{3}}|-\frac{1}{2}\rangle + \alpha_{1}\alpha_{2}\alpha_{3}|-\frac{3}{2}\rangle $\\

      $\rightarrow$  $[1,\frac{(\alpha_{1}+\alpha_{1}+\alpha_{1})}{\sqrt{3}},\frac{(\alpha_{1}\alpha_{2}+\alpha_{2}\alpha_{3}+\alpha_{3}\alpha_{1})}{\sqrt{3}},\alpha_{1}\alpha_{2}\alpha_{3}]$ \quad \quad \quad (2)\\

\noindent
where $|m\rangle$ with $m = +\frac{3}{2},+\frac{1}{2},-\frac{1}{2}$ or $-\frac{3}{2}$ are spin eigenstates along the z-axis in the standard angular momentum basis. The second line of (2) shows the state more briefly as a four-component column vector (both here and later all such vectors are to be interpreted as column vectors, although they are displayed as row vectors for convenience).\\

The expressions for the explicit rays can be worked out from (1) and (2) by noting that two of the alpha parameters of any explicit ray are those of the vertex representing it while the third is that of the antipodal vertex. One then finds that the vectors representing the normalized explicit rays are\\

$ A = [0,1,0,0],  \quad  T = [0,0,1,0] $ \\[1pt]

$F = [\frac{\tau}{3},-\frac{1}{\surd3},-\frac{1}{\surd3},-\frac{1}{3\tau}], \quad B = [\frac{\tau}{3},-\frac{\omega}{\surd3},-\frac{\omega^{2}}{\surd3},-\frac{1}{3\tau}], \quad E = [\frac{\tau}{3},-\frac{\omega^{2}}{\surd3},-\frac{\omega}{\surd3},-\frac{1}{3\tau}]$ \\[1pt]

$L = [\frac{2}{3},0,-\frac{1}{4}(\frac{1}{\sqrt{3}}+i\sqrt{5}),\frac{1}{4}(\frac{\sqrt{5}}{3}-i\sqrt{3})]$ \\[1pt]

$K = [\frac{2}{3},0,-\frac{1}{4}(\frac{1}{\sqrt{3}}-i\sqrt{5}),\frac{1}{4}(\frac{\sqrt{5}}{3}+i\sqrt{3})]$ \\[1pt]

$G = [\frac{2}{3},0,\frac{\sqrt{3}}{8}(\frac{1}{3}+\sqrt{5})+\frac{i}{8}(1-\sqrt{5}),\frac{1}{4}(\frac{\sqrt{5}}{3}+i\sqrt{3})]$ \\[1pt]

$J = [\frac{2}{3},0,\frac{\sqrt{3}}{8}(\frac{1}{3}+\sqrt{5})-\frac{i}{8}(1-\sqrt{5}),\frac{1}{4}(\frac{\sqrt{5}}{3}-i\sqrt{3})]$ \\[1pt]

$C = [\frac{2}{3},0,\frac{\sqrt{3}}{8}(\frac{1}{3}-\sqrt{5})+\frac{i}{8}(1+\sqrt{5}),\frac{1}{4}(\frac{\sqrt{5}}{3}-i\sqrt{3})]$ \\[1pt]

$D = [\frac{2}{3},0,\frac{\sqrt{3}}{8}(\frac{1}{3}-\sqrt{5})-\frac{i}{8}(1+\sqrt{5}),\frac{1}{4}(\frac{\sqrt{5}}{3}+i\sqrt{3})]$ \\[1pt]

$N = [\frac{1}{3\tau},-\frac{1}{\surd3},\frac{1}{\surd3},\frac{\tau}{3}], \quad U = [\frac{1}{3\tau},-\frac{\omega}{\surd3},\frac{\omega^{2}}{\surd3},\frac{\tau}{3}], \quad S = [\frac{1}{3\tau},-\frac{\omega^{2}}{\surd3},\frac{\omega}{\surd3},\frac{\tau}{3}]$ \\[1pt]

$I = [\frac{\sqrt{2}}{3},\frac{1}{2\sqrt{2}}(-\sqrt{\frac{5}{3}}+i),0,-\frac{1}{2\sqrt{2}}(\frac{\sqrt{5}}{3}-i\sqrt{3})] $ \\[1pt]

$H = [\frac{\sqrt{2}}{3},\frac{1}{2\sqrt{2}}(-\sqrt{\frac{5}{3}}+i),0,-\frac{1}{2\sqrt{2}}(\frac{\sqrt{5}}{3}+i\sqrt{3})] $ \\[1pt]

$P = [\frac{\sqrt{2}}{3},\frac{1}{4\sqrt{2}}(\sqrt{\frac{5}{3}}-\sqrt{3})-i(\sqrt{5}+1),0,-\frac{1}{2\sqrt{2}}(\frac{\sqrt{5}}{3}+i\sqrt{3})] $ \\[1pt]

$M = [\frac{\sqrt{2}}{3},\frac{1}{4\sqrt{2}}(\sqrt{\frac{5}{3}}-\sqrt{3})+i(\sqrt{5}+1),0,-\frac{1}{2\sqrt{2}}(\frac{\sqrt{5}}{3}-i\sqrt{3})] $ \\[1pt]

$Q = [\frac{\sqrt{2}}{3},\frac{1}{4\sqrt{2}}(\sqrt{\frac{5}{3}}+\sqrt{3})-i(\sqrt{5}-1),0,-\frac{1}{2\sqrt{2}}(\frac{\sqrt{5}}{3}-i\sqrt{3})] $ \\[1pt]

$R = [\frac{\sqrt{2}}{3},\frac{1}{4\sqrt{2}}(\sqrt{\frac{5}{3}}+\sqrt{3})+i(\sqrt{5}-1),0,-\frac{1}{2\sqrt{2}}(\frac{\sqrt{5}}{3}+i\sqrt{3})] $ \quad , \quad (3)\

\noindent\\
where $\tau = \frac{1+\surd5}{2}$ is the golden ratio and $\omega = e^{2\pi i/3}$ is a cube root of unity. The (normalized) implicit ray $A'$ orthogonal to the explicit rays $F$,$E$ and $B$ is easily worked out to be\footnote{The extra factor of $i$ in the components is unnecessary but has been added for later convenience.}\\

$ \quad \quad \quad \quad A' = [\frac{i}{\sqrt{3}\tau},0,0,\frac{i\tau}{\sqrt{3}}] $ \quad . \quad (4)\\

\noindent\\
The expressions for the explicit rays can be put in a simpler form by switching from the angular momentum basis to a ``natural'' basis consisting of the mutually orthogonal rays $F$, $B$, $E$ and $A'$.\footnote{It was pointed out in \cite{Massad} that switching to this basis would simplify the expressions for the Penrose rays, but the connection with the Witting polytope was not realized.} This change can be effected by multiplying the (column) vectors of the explicit rays in (3) by the unitary matrix \\

\quad \quad \quad
$\Omega =
\left( \begin{array}{cccc}
\frac{\tau}{3} & -\frac{1}{\sqrt{3}} & -\frac{1}{\sqrt{3}} & -\frac{1}{3\tau} \\
\frac{\tau}{3} & -\frac{\omega}{\sqrt{3}} & -\frac{\omega^{2}}{\sqrt{3}} & -\frac{1}{3\tau} \\
\frac{\tau}{3} & -\frac{\omega^{2}}{\sqrt{3}} & -\frac{\omega}{\sqrt{3}} & -\frac{1}{3\tau} \\
-\frac{i}{\sqrt{3}\tau} & 0 & 0 & -\frac{i\tau}{\sqrt{3}} \end{array} \right)$   \quad . \quad (5) \\
\\

\noindent\\
which takes the rays $F$, $B$, $E$ and $A'$ into the column vectors $[1,0,0,0]$, $[0,1,0,0]$, $[0,0,1,0]$ and $[0,0,0,1]$, respectively, but yields somewhat complicated expressions for the other rays. However if the other rays are rescaled so that their leading nonvanishing components are unity, they take on the simple forms shown in Table 1. The forms of the implicit rays in this basis are easily worked out by making them orthogonal to the appropriate triads of explicit rays, and we have also rescaled them to make their leading nonvanishing components unity. Table 1 shows the Penrose rays in their simplest possible form.

\begin{table}[ht]
\centering % used for centering table
\begin{tabular}{|c | c | c | c |} % centered columns (4 columns)
\hline
  $ $ & $ $ & $ $ & $ $\\
  $F = [1000]$   & $B = [0100]$ & $E = [0010]$ & $A' = [0001]$  \\
  $ $ & $ $ & $ $ & $ $\\
  $N = [01\bar{1}1]$   & $U = [10\bar{1}\bar{1}]$ & $S = [1\bar{1}01]$ & $T' = [1110]$  \\
  $ $ & $ $ & $ $ & $ $\\
  $Q = [01\bar{\omega}1]$   & $L = [10\bar{\omega}\bar{1}]$ & $I = [1\bar{\omega}01]$ & $P' = [1\omega10]$  \\
  $ $ & $ $ & $ $ & $ $\\
  $G = [01\bar{\omega^{2}}1]$   & $H = [10\bar{\omega^{2}}\bar{1}]$ & $K = [1\bar{\omega^{2}}01]$ & $Q' = [1\omega^{2}10]$  \\
  $ $ & $ $ & $ $ & $ $\\
  $K' = [01\bar{1}\omega]$   & $G' = [10\bar{1}\bar{\omega}]$ & $D' = [1\bar{1}0\omega]$ & $R' = [11\omega0]$  \\
  $ $ & $ $ & $ $ & $ $\\
  $R = [01\bar{\omega}\omega]$   & $D = [10\bar{\omega}\bar{\omega}]$ & $P = [1\bar{\omega}0\omega]$ & $I' = [1\omega\omega0]$  \\
  $ $ & $ $ & $ $ & $ $\\
  $F' = [01\bar{\omega^{2}}\omega]$   & $U' = [10\bar{\omega^{2}}\bar{\omega}]$ & $E' = [1\bar{\omega^{2}}0\omega]$ & $A = [1\omega^{2}\omega0]$  \\
  $ $ & $ $ & $ $ & $ $\\
  $L' = [01\bar{1}\omega^{2}]$   & $C' = [10\bar{1}\bar{\omega^{2}}]$ & $J' = [1\bar{1}0\omega^{2}]$ & $M' = [11\omega^{2}0]$  \\
  $ $ & $ $ & $ $ & $ $\\
  $N' = [01\bar{\omega}\omega^{2}]$   & $B' = [10\bar{\omega}\bar{\omega^{2}}]$ & $S' = [1\bar{\omega}0\omega^{2}]$ & $T = [1\omega\omega^{2}0]$  \\
  $ $ & $ $ & $ $ & $ $\\
  $J = [01\bar{\omega^{2}}\omega^{2}]$   & $M = [10\bar{\omega^{2}}\bar{\omega^{2}}]$ & $C = [1\bar{\omega^{2}}0\omega^{2}]$ & $H' = [1\omega^{2}\omega^{2}0]$\\
  $ $ & $ $ & $ $ & $ $\\
\hline
\end{tabular}
\caption{The 40 rays of the Penrose dodecahedron. Unprimed letters represent the explicit rays and primed ones the implicit rays. The rays in the first row define the basis and therefore have the simple forms shown; with this choice, the rest of the rays assume the forms shown when they are rescaled to make their leading nonvanishing components unity. The rays in each column (with the exception of those in the first row) have a 0 in the same place and all the nonvanishing components of the rays are signed cube roots of unity. Note: commas have been omitted between the components and a bar over an entry denotes its negative.}
\label{table1} % is used to refer this table in the text
\end{table}

\section{\label{sec:3} The Witting polytope}

Coxeter defines the notion of a regular complex polytope in his book of that name \cite{Coxeter} and provides many examples of such objects. Briefly, a complex polytope is an object in a complex unitary space of $n$ dimensions, $\mathbb{C}^{n}$, whose vertices each have $n$ complex coordinates and whose edges can each have more than two vertices associated with them. A ``$p$-edge'' of a complex polytope has $p$ vertices associated with it, each of which can be made to pass into a neighboring vertex by a rotation of $2\pi/p$ about the center of the edge\footnote{Thus a real polytope may be regarded as consisting entirely of 2-edges.}. The Witting polytope is an object in $\mathbb{C}^{4}$ with 240 vertices and 2160 3-edges. It has many interesting properties that are discussed at length in an article by Coxeter and Shephard \cite{leonardo}, which also shows a highly symmetrical two-dimensional projection of it obtained using a simple technique. All that one needs to know about the Witting polytope, for the purposes of this paper, is that it has 240 vertices with the complex coordinates shown in Table 2.

\begin{table}[ht]
\centering % used for centering table
\begin{tabular}{|c | c | c | c |} % centered columns (4 columns)
\hline
$ $ & $ $ & $ $ & $ $\\
  $[0,\pm\omega^{\mu},\mp\omega^{\nu},\pm\omega^{\lambda}]$ & $[\mp\omega^{\mu},0,\pm\omega^{\nu},\pm\omega^{\lambda}]$ & $[\pm\omega^{\mu},\mp\omega^{\nu},0,\pm\omega^{\lambda}]$ & $[0,\mp\omega^{\mu},\mp\omega^{\nu},\mp\omega^{\lambda}]$ \\
$ $ & $ $ & $ $ & $ $\\
$[\pm i\omega^{\lambda\surd3},0,0,0]$ & $[0,\pm i\omega^{\lambda\surd3},0,0]$ & $[0,0,\pm i\omega^{\lambda\surd3},0]$ & $[0,0,0,\pm i\omega^{\lambda\surd3}]$ \\
$ $ & $ $ & $ $ & $ $\\
\hline
\end{tabular}
\caption{Coordinates of the 240 vertices of the Witting polytope, with $\omega = \exp(2\pi i/3)$ and $\mu, \nu$ and $\lambda$ each being allowed to take on the values 0,1 and 2 independently. Each of the entries in the first row represents 54 vertices while each in the second row represents 6 vertices.}
\label{table1} % is used to refer this table in the text
\end{table}

The vertices of the Witting polytope, which are elements of $\mathbb{C}^{4}$, can be mapped into rays in $\mathbb{CP}^{3}$, with all vertices having coordinates that differ by an overall constant being mapped into the same ray. This leads to a 6 to 1 mapping of vertices into rays, with each block of vertices in the first row of Table 2 giving rise to 9 rays and each block in the second row to a single ray. If one rescales all the rays so that their leading non-vanishing components are unity, one gets back just the 40 rays listed in Table 1. The unitary equivalence of the rays of the Witting polytope to those of the Penrose dodecahedron is thereby established.

\section{\label{sec:4}Discussion}

The 40 rays of the Penrose dodecahedron form the 40 bases (i.e., sets of four mutually orthogonal rays) shown in Table 3. These bases were used by Zimba and Penrose \cite{Zimba} to give non-coloring proofs of the KS and Bell theorems (an alternative version of these proofs can also be found in \cite{Massad}). \\

\begin{table}[ht]
\centering % used for centering table
\begin{tabular}{|c | c | c | c |} % centered columns (4 columns)
\hline
  $FBEA'$ & $LQSR'$ & $AA'TT'$ & $A'I'Q'M'$  \\
  $ALKF'$ & $GHSM'$ & $BB'UU'$ & $A'P'R'H'$ \\
  $AGCB'$ & $CMNH'$ & $CC'QQ'$ & $B'K'I'S'$ \\
  $ADJE'$ & $DPNI'$ & $DD'RR'$ & $B'J'N'R'$ \\
  $FGRL'$ & $JIUP'$ & $EE'SS'$ & $C'F'P'S'$ \\
  $BLMG'$ & $KRUQ'$ & $FF'NN'$ & $C'L'J'T'$ \\
  $BDHC'$ & $RMTS'$ & $GG'PP'$ & $D'K'G'T'$ \\
  $ECID'$ & $HITN'$ & $HH'KK'$ & $D'F'M'U'$ \\
  $EKPJ'$ & $PQTU'$ & $II'LL'$ & $E'L'H'U'$ \\
  $FJQK'$ & $SNUT'$ & $JJ'MM'$ & $E'G'N'Q'$ \\
\hline
\end{tabular}
\caption{The 40 bases formed by the 40 rays of the Penrose dodecahedron. The bases in the first two columns each involve the implicit ray associated with a vertex of the dodecahedron and the explicit rays associated with the three neighboring vertices; the bases in the third column involve the explicit and implicit rays associated with a pair of antipodal vertices; and the bases in the last column involve the implicit rays associated with the vertices of the ten tetrahedra that can be inscribed in the dodecahedron.}
\label{table1} % is used to refer this table in the text
\end{table}

The main contribution of this paper is to dispel the aura of mystery surrounding the original introduction of the Penrose rays by pointing out an alternative, and simpler, method by which they can be arrived at: one need only look at the system of rays associated with the vertices of the Witting polytope, whereupon both the expressions for the simplified Penrose rays (Table 1) and their basis table (Table 3) follow immediately. With these in hand, one can proceed to the proofs of the KS and Bell theorems as before. It is interesting to realize, after all these years, that the Penrose dodecahedron already existed within the bowels of the Witting polytope, so to speak, before Penrose came up with his ingenious construction of it.\\

Coxeter and Shephard \cite{leonardo} point out that any complex polytope in $\mathbb{C}^{n}$ has a real representative in $\mathbb{R}^{2n}$ with the same number of vertices, but with each vertex having twice as many coordinates, these being the real and imaginary parts of the coordinates of the complex polytope. The Witting polytope gives rise to a real polytope in $\mathbb{R}^{8}$ with 240 vertices known as Gosset's polytope\footnote{This polytope is described by the symbol $4_{21}$, and is not to be confused with the six- and seven-dimensional polytopes also named after Gosset. See Coxeter \cite{Cox} for a fuller discussion of these objects.}, with the property that the vectors from its center to its vertices are the root vectors of the exceptional Lie algebra E8. Viewed in $\mathbb{RP}^{7}$ (i.e., a real projective space of seven dimensions), the 240 vertices of Gosset's polytope collapse into 120 rays that form 2025 bases of eight mutually orthogonal rays. This system of rays and bases can be characterized by the symbol $120_{135}-2025_{8}$, with the subscript on the left indicating that each of the rays occurs in 135 bases and that on the right that each basis consists of 8 rays (note that $120 \times 135 = 2025 \times 8$, as it should be). In a recent paper \cite{E8} we showed that this $120_{135}-2025_{8}$ system of rays and bases provides a large number of parity proofs of the KS theorem that take no more than simple counting to verify (unlike the much more involved proofs provided by the Penrose dodecahedron). Thus the Witting polytope is remarkable in that both it and its real representative (the Gosset polytope $4_{21}$) provide proofs of the KS theorem, but of very different kinds. We are not aware of any other complex polytopes that exhibit this dual (almost chameleon-like?) behavior, and it would be interesting to identify them if they do exist.\\

The Witting polytope, in its avatar as the Penrose dodecahedron, is not the only source of a $40_{4}-40_{4}$ system of rays and bases that provides proofs of the KS theorem. In \cite{Waegell2011c} we presented another $40_{4}-40_{4}$ system, derived from the eigenstates of commuting sets of observables of a two-qubit system, that also provides such proofs. However these two systems differ in a number of respects. Firstly, their Kochen-Specker diagrams are different; this is evident from the fact that each ray in the Penrose-Witting system is orthogonal to 12 others and each in the two-qubit system to 9 others (this also implies that the two systems are not unitarily equivalent). And, secondly, the bases in the two systems have a very different character. The bases of the Penrose-Witting system are closely tied to the geometry of a dodecahedron, as explained in the caption to Table 3. By contrast, the bases of the two-qubit system can be divided into ten ``pure`` bases, obtained as the simultaneous eigenstates of ten triads of commuting observables, and thirty ``hybrid`` bases, each made up of pairs of states from a pair of pure bases (see Fig.1 of \cite{Waegell2011c} for a diagram showing the interrelationship between the pure and hybrid bases). The consequence of this difference in basis structure is that the two-qubit system has $2^{15}=32768$ parity proofs in it, whereas the Penrose-Witting system has not a single proof of this kind (although it does have more involved proofs instead).\\

The two-qubit system actually gives rise to a $60_{7}-105_{4}$ system of rays and bases if one considers the simultaneous eigenstates of all 15 triads of commuting observables in it and constructs all the bases formed by these states. This larger system contains the $40_{4}-40_{4}$ system discussed earlier as a subset (and actually has six such systems in it, all partially overlapping each other). The larger system has identical parity proofs in each of its six subsystems, but it also has new proofs not contained in any of its subsystems. This leads one to ask if the Penrose-Witting system might be embedded in a larger structure that contains other systems like it, and that yields new proofs not contained in any of the subsystems. We now indicate how such a larger structure can be constructed. Consider the rays of the computational basis, $[1,0,0,0], [0,1,0,0], [0,0,1,0]$ and $[0,0,0,1]$ along with the rays\\
\noindent

$ F_{1}(i,j,k,l) = [0,1,(-1)^{i}\omega^{k},(-1)^{j}\omega^{l}]$ \\

$ F_{2}(i,j,k,l) = [1,0,(-1)^{i}\omega^{k},(-1)^{j}\omega^{l}]$ \\

$ F_{3}(i,j,k,l) = [1,(-1)^{i}\omega^{k},0,(-1)^{j}\omega^{l}]$ \\

$ F_{4}(i,j,k,l) = [1,(-1)^{i}\omega^{k},(-1)^{j}\omega^{l},0]$ \quad , \quad (6) \\

\noindent
where $i,j = 0,1$ and $k,l = 0,1,2$, so that each line represents 36 rays. These 148 rays form 265 bases and can be characterized by the symbol $4_{13}144_{7}-265_{4}$, which indicates that 4 rays each occur 13 times and 144 rays 7 times among the 265 bases of the system. This system has eight $40_{4}-40_{4}$ subsystems of the Penrose-Witting type in it. Each subsystem consists of the four rays of the computational basis and the 36 rays in one of the following eight lines:\\
\noindent

$F_{1}(0,0,i,j)$ , \quad $F_{2}(0,1,i,j)$ , \quad $F_{3}(1,0,i,j)$,  \quad $F_{4}(0,1,i,j)$ \\

$F_{1}(0,0,i,j)$ , \quad $F_{2}(1,0,i,j)$ , \quad $F_{3}(0,1,i,j)$,  \quad $F_{4}(1,0,i,j)$ \\

$F_{1}(0,1,i,j)$ , \quad $F_{2}(0,0,i,j)$ , \quad $F_{3}(1,1,i,j)$,  \quad $F_{4}(0,1,i,j)$ \\

$F_{1}(0,1,i,j)$ , \quad $F_{2}(1,1,i,j)$ , \quad $F_{3}(0,0,i,j)$,  \quad $F_{4}(1,0,i,j)$ \\

$F_{1}(1,0,i,j)$ , \quad $F_{2}(0,0,i,j)$ , \quad $F_{3}(0,1,i,j)$,  \quad $F_{4}(1,1,i,j)$ \\

$F_{1}(1,0,i,j)$ , \quad $F_{2}(1,1,i,j)$ , \quad $F_{3}(1,0,i,j)$,  \quad $F_{4}(0,0,i,j)$ \\

$F_{1}(1,1,i,j)$ , \quad $F_{2}(0,1,i,j)$ , \quad $F_{3}(0,0,i,j)$,  \quad $F_{4}(1,1,i,j)$ \\

$F_{1}(1,1,i,j)$ , \quad $F_{2}(1,0,i,j)$ , \quad $F_{3}(1,1,i,j)$,  \quad $F_{4}(0,0,i,j)$ \quad (7) \\

\noindent
These eight $40_{4}-40_{4}$ systems are unitarily equivalent to each other. This can be seen by noting that a permutation of the rays of the computational basis, followed by a suitable rephasing, can be used to turn one of the systems into another.\footnote{For example, the basis permutation
$[1,0,0,0] \rightarrow -[0,1,0,0], [0,1,0,0] \rightarrow [0,0,0,1], [0,0,1,0] \rightarrow -[1,0,0,0]$ and $[0,0,0,1] \rightarrow [0,0,1,0]$ followed by a suitable rephasing of the rays results in the transformation $F_{1}(0,0)\rightarrow F_{2}(1,1), F_{2}(0,1) \rightarrow F_{4}(0,0), F_{3}(1,0) \rightarrow F_{1}(1,0)$ and $F_{4}(0,1) \rightarrow F_{3}(1,0)$, which causes the system described in the first line of (7) to go into that described in the sixth.} The open question is whether this enlarged system of 148 rays and 265 bases contains new proofs of the KS theorem not contained in any of the $40_{4}-40_{4}$ subsystems in it. We suspect this to be the case, but this is a point that still remains to be established. \\

It might be mentioned in closing that the Witting polytope turns up in yet another problem in quantum mechanics, namely, that of constructing projective t-designs \cite{Hoggar}. t-designs are closely related to SIC-POVMs (Symmetric Informationally Complete Positive Operator Valued Measures), which have been studied extensively by physicists in connection with the foundations of quantum mechanics and their application to quantum state tomography \cite{Fuchs,Caves2004,Scott,Stacey}. \\

A summary of the main points of this paper is given in Fig.2 and the caption below it.

\begin{figure}[htp]
\begin{center}
\includegraphics[width=0.60\textwidth]{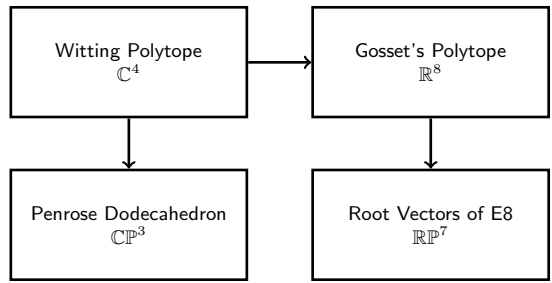}
\end{center}
\caption{The Witting polytope, top left, is a geometrical figure in $\mathbb{C}^{4}$. The horizontal arrow to its right leads to its real representative in $\mathbb{R}^{8}$, Gosset's polytope $4_{21}$. The downward arrow from the Witting polytope indicates that as one descends from $\mathbb{C}^{4}$ into $\mathbb{CP}^{3}$ one obtains the system of rays associated with the Penrose dodecahedron; this is the main message of this paper. The downward arrow at the right indicates that as one descends from $\mathbb{R}^{8}$ into $\mathbb{RP}^{7}$, Gosset's polytope yields a system of rays associated with the root vectors of the Lie algebra E8. While the two boxes at the top contain geometric objects, the boxes below them contain derived systems of rays that are useful in proofs of the KS and Bell theorems. Proofs based on the Penrose dodecahedron were given in \cite{Zimba}, while proofs based on the rays of E8 were given in \cite{E8}. The Witting polytope might be termed a ``quantum chameleon'' because it leads, via different routes, to the two boxes at the bottom that provide very different proofs of the KS and Bell theorems.}
\label{summary}
\end{figure}

%\bibliographystyle{ieeetr}
%\bibliography{testbib}

\end{document}